\documentclass[conference,a4]{IEEEtran}

\usepackage[utf8]{inputenc} 
\usepackage[T1]{fontenc}
\usepackage{url}
\usepackage{ifthen}
\usepackage{cite}
\usepackage[cmex10]{amsmath} %
\usepackage{amsfonts}
\usepackage{bm}
\usepackage{acronym}
\newacro{4SET}[4SET]{four degree-one VN stopping set}

\newacro{AU-BAC}[AU\nobreakdash-BAC]{asynchronous U\nobreakdash-BAC}

\newacro{BEC}[BEC]{binary erasure channel}
\newacro{BPSK}[BPSK]{binary phase shift keying}

\newacro{CN}[CN]{check node}

\newacro{DE}[DE]{density evolution}

\newacro{LDPC}[LDPC]{low-density parity-check}
\newacroindefinite{LDPC}{an}{a}

\newacro{mMTC}[mMTC]{massive machine-type communication}
\newacro{MN}[MN]{multiple access node}
\newacroindefinite{MN}{an}{a}

\newacro{PCM}[PCM]{parity-check matrix}
\newacro{PEG}[PEG]{progressive edge growth}
\newacro{PUPE}[PUPE]{per user probability of error}

\newacro{RCC}[RCC]{random code construction}

\newacro{U-BAC}[U\nobreakdash-BAC]{unsourced two-user binary adder channel}
\newacro{UMAC}[UMAC]{unsourced multiple access channel}

\newacro{VN}[VN]{variable node}

\usepackage{tikz}
\usepackage{pgf, tikz, pgfplots}
\usetikzlibrary{pgfplots.fillbetween,shapes,patterns,patterns.meta,calc,positioning}
\pgfplotsset{compat=1.18}
\usepackage{algorithm}
\usepackage[noend]{algorithmic}
\usepackage{mathtools}
\usepackage{aligned-overset}
\usepackage{siunitx}
\usepackage{booktabs}
\usepackage{cite}

\usepackage{kitcolors}

\newtheorem{theorem}{Theorem}

\newcommand\blfootnote[1]{%
  \begingroup
  \renewcommand\thefootnote{}\footnote{#1}%
  \addtocounter{footnote}{-1}%
  \endgroup
}

\newcounter{enscounter}\renewcommand{\theenscounter}{\arabic{enscounter}}
\newcommand{\enslabel}[1]{%
  \refstepcounter{enscounter}\theenscounter\label{#1}%
}

\begin{document}

\title{Removal of Small Weight Stopping Sets for Asynchronous Unsourced Multiple Access}

\author{%
  \IEEEauthorblockN{%
    Frederik Ritter\textsuperscript{\dag},
    Jonathan Mandelbaum\textsuperscript{\dag},
    Alexander Fengler\textsuperscript{\ddag},
    Holger Jäkel\textsuperscript{\dag},
    and Laurent Schmalen\textsuperscript{\dag}
  }
  \IEEEauthorblockA{%
    \textsuperscript{\dag}Karlsruhe Institute of Technology (KIT), Communications Engineering Lab (CEL), 76187 Karlsruhe, Germany\\
    \textsuperscript{\ddag}German Aerospace Center (DLR) Oberpfaffenhofen, 82234 Weßling, Germany\\
    Email: {\texttt{\{frederik.ritter, jonathan.mandelbaum\}@kit.edu}
  }
}

\vspace{-1em}}

\maketitle

\begin{abstract}
  In this paper, we analyze the formation of small stopping sets in joint factor graphs describing a frame-asynchronous two-user transmission.
  Furthermore, we propose an algorithm to completely avoid small stopping sets in the joint factor graph over the entire range of symbol delays.
  The error floor caused by these stopping sets is completely mitigated.
  Our key observation is that, while the order of bits in the codeword is irrelevant in a single-user environment, it turns out to be crucial in an asynchronous, unsourced two-user system.
  Subsequently, our algorithm finds a reordering of variable nodes which avoids the smallest stopping set in the joint graph.
  We show that further improvements can be achieved when girth optimization of the single-user graphs by progressive edge growth (PEG) is used in combination with our proposed algorithm.
  Starting with a randomized code construction with optimized degree distribution, our simulation results show that PEG followed by the proposed algorithm can improve the average per user probability of error in a noiseless channel by almost two orders of magnitude for a broad range of frame delays.
\end{abstract}

\blfootnote{This work has received funding from the
German Federal Ministry of Education and Research (BMBF) within the project Open6GHub (grant agreement 16KISK010) and the European Research Council (ERC) under the European Union’s Horizon 2020 research and innovation programme (grant agreement No. 101001899).}
\section{Introduction}

Grant-based multiple access schemes are facing a new challenge with the advent of \ac{mMTC}~\cite{choi_grant-free_2022,yuan_non-orthogonal_2016}.
As many transmitters are sending short payloads at random intervals, the signaling required for grant acquisition constitutes a considerable, non-negligible fraction of all transmissions.
This reduces the efficiency of the system, taking a toll on, e.g., the energy consumption of the devices.

One contender for a more efficient approach is the \ac{UMAC}~\cite{polyanskiy_perspective_2017}.
In a \ac{UMAC}, all transmitters use the same code and the receiver only recovers the list of transmitted codewords, without attributing codewords to individual transmitters.
In this unsourced scheme, user identification is assumed to be handled in another protocol layer, e.g., using headers.
Thus, user identification and combating interference on a \ac{UMAC} are separate tasks that can be tackled individually.
For more information on the \ac{UMAC} and its relevance for next-generation communication standards, the interested reader is referred to~\cite{miao_trends_2024,liva_unsourced_2024}.

In~\cite{liva_coding_2021}, Liva \emph{et al.} study the \ac{U-BAC}, which is a simplified version of the \ac{UMAC} in the form of a synchronous channel with only two users and no noise.
Destructive interference occurs on this channel if the two users transmit opposing symbols.
This can be modelled as an erasure, indicating similarities to the \ac{BEC} and enabling the use of \acf{LDPC} codes and peeling decoding to combat interference between the two users.
A challenge described by Liva \emph{et al.} is that only rates below $\frac{1}{2}$ bit/channel use are achievable using linear codes, while the per-user capacity of their channel model is $\frac{3}{4}$ bit/channel use.
In~\cite{fengler_advantages_2023}, it is shown that this
limitation 
for linear codes can be overcome by including frame asynchrony in the channel model.
Additionally, a program to find optimized \ac{LDPC} code ensembles is derived in~\cite{fengler_advantages_2023}.
Hereby, the error probability is shown to tend to zero if the rate $r < \frac{3}{4}$ and the block length $n \to \infty$, even if a small fraction of degree-one \acfp{VN} is present.
The advantage of introducing degree-one \acp{VN} is that they reduce the average \ac{VN} degree $d_\mathrm{v,avg}$ and, therefore, allow to potentially increase the design rate $r_\mathrm{d} = 1 - \frac{d_\mathrm{v,avg}}{d_\mathrm{c,avg}}$~\cite{richardson_modern_2008}, where $d_\mathrm{c,avg}$ denotes the average \ac{CN} degree.
However, these degree-one \acp{VN} can also lead to small stopping sets which result in a high error floor.
In~\cite{fengler_advantages_2023} it is shown that small stopping sets are unlikely to occur in a random construction for a fixed, non-zero frame delay $\tau$, yet they are highly likely to occur if multiple different values of $\tau$ are considered.

Here, we study code constructions that eliminate small weight stopping sets for all $\tau \geq 1$.
To this end, we analyze how the smallest degree-one \ac{VN} stopping sets form.
In contrast to a single-user code, where the ordering of codebits is irrelevant to the decoding performance, we observe that the order of columns of a \ac{PCM} influences the multi-user performance of the code.
Hence, we propose an algorithm to remove the smallest degree-one \ac{VN} stopping sets by simply permuting columns of a given \ac{PCM}.
Furthermore, we find that removing these stopping sets is simplified if an approximately equal number of degree-one \acp{VN} is connected to each \ac{CN}.

\section{Preliminaries} \label{sec:preliminaries}

In the following, we study the \ac{U-BAC}, which is defined as
\begin{equation}
  Y_i = X^{(1)}_i + X^{(2)}_i,
\end{equation}
where $X^{(1)}_i, X^{(2)}_i \in \{-1, 1\}$ are the \ac{BPSK} symbols sent by the two transmitters at time step ${i \in [n]:=\{1,\dots,n\}}$ and $Y_i \in \{-2, 0, 2\}$ is the received value.
Both codewords $\bm{X}^{(1)}, \bm{X}^{(2)} \in \{-1, 1\}^n$ stem from the same code.
Although this \ac{UMAC} is noise-free, we expect that its study is useful for channels with noise.
For instance, in~\cite{liva_coding_2021}, the error floor of a code in a noisy two-user channel is explained by its performance on the \ac{U-BAC}.

\ac{LDPC} codes are binary linear block codes defined by the null space of a sparse \acf{PCM} ${\bm{H}\in \mathbb{F}_2^{m\times n}}$~\cite{richardson_modern_2008}.
The associated \emph{Tanner graph} is a bipartite graph consisting of \acp{VN} and \acp{CN}, where each \ac{VN} represents a column and each \ac{CN} represents a row of $\bm{H}$, respectively.
A \ac{CN}~$c_j$ is connected to a \ac{VN}~$v_i$ if ${H_{ji} = 1}$, where $H_{ji}$ is the entry of $\bm{H}$ at row $j$ and column $i$.
The degree of a node in the Tanner graph is the number of edges incident to it.
The fraction of degree\nobreakdash-$i$ \acp{VN} is denoted by $L_i$ and the fraction of degree\nobreakdash-$j$ \acp{CN} is denoted by $R_j$, respectively.
We define the tuples $(L_1, L_2, \ldots,L_{d_\mathrm{v,max}})$ and $(R_1, R_2, \ldots,R_{d_\mathrm{c,max}})$ as the \ac{VN} and \ac{CN} degree distributions, respectively.
Here, $d_\mathrm{v,max}$ and $d_\mathrm{c,max}$ denote the maximum \ac{VN} and \ac{CN} degree, respectively.
All codes with a given pair of \ac{VN} and \ac{CN} degree distributions constitute an \ac{LDPC} code \emph{ensemble}.

While, for general (non-linear) codes, the per-user capacity of the \ac{U-BAC} is $\frac{3}{4}$ bit/channel use, linear codes, including \ac{LDPC} codes, can only achieve a rate of $\frac{1}{2}$ bit/channel use~\cite{liva_coding_2021}.
\emph{Mild asynchrony} is a form of asynchrony where the delay between the two transmitters grows sub-linearly with the block length.
Thus, the fraction of non-overlapping symbols becomes negligible as the block length $n \to \infty$ and the capacity region of the channel remains unaffected~\cite{el_gamal_network_2011}.
Allowing for mild asynchrony, i.e., a delay $o(n) > \tau \geq 1$ between the two transmitters, increases the achievable rate using linear codes to the $\frac{3}{4}$ bit/channel use~\cite{fengler_advantages_2023}.
Additionally, a coding scheme tolerant to asynchrony simplifies the system design by relaxing the requirements for frame synchronization~\cite{fengler_advantages_2023}.
Thus, following~\cite{fengler_advantages_2023}, we extend the \ac{U-BAC} to the \ac{AU-BAC} by introducing an arbitrary but fixed delay $\tau \in [\tau_\text{max}]$ such that
\begin{equation}
  Y_i = X^{(1)}_i + X^{(2)}_{i-\tau}.
\end{equation}
In contrast to the \ac{U-BAC}, the received value $Y_i$ is defined for $i\in[n+\tau]$, and, by using the convention that ${X^{(1)}_i = X^{(2)}_i = 0}$ if $i\notin[n]$, we find $Y_i \in \{-1, 1\}$ if $i \in [\tau] \cup \{n+1, \dots, n+\tau\}$, while $Y_i \in \{-2, 0, 2\}$ otherwise.
This change in the output alphabet enables the receiver to unambiguously retrieve the delay $\tau$.
While mild asynchrony is required for the asymptotic analysis of the achievable rate, it imposes an additional constraint on real-world implementations.
Thus, we consider the more general case of $n \geq \tau \geq 1$ in the remaining paper.

\begin{figure}[tbp]
  \centering
  \begin{tikzpicture}[%
  every node/.style={fill=gray, minimum size = 2.5mm, inner sep = 0},
  vn/.style={draw, circle},
  cn/.style={draw, rectangle},
  mac/.style={
    draw,
    color=KITgreen,
    isosceles triangle,
    rotate = 90,
    isosceles triangle apex angle = 60,
  },
  label/.style={fill=none, anchor = east}
]

  \node[label] at (.3,0) {User $1$ CNs};
  \node[label] at (.3,-1*0.8) {User $1$ VNs};
  \node[label, color=KITgreen] at (.3,-1.5*0.8) {MNs};
  \node[label] at (.3,-2*0.8) {User $2$ VNs};
  \node[label] at (.3,-3*0.8) {User $2$ CNs};

  \foreach \i in {1,2,3}
    \node[cn] (cn1\i) at (2*0.8*\i,0) {};

  \foreach \i in {1,2,...,7}
    \node[vn] (vn1\i) at (0.8*\i, -1*0.8) {};

  \foreach \i in {2,3,...,7}
    \node[mac] (mac\i) at (0.8*\i, -1.5*0.8) {};

  \foreach \i in {1,2,...,7}
    \node[vn] (vn2\i) at (0.8*\i + 0.8, -2*0.8) {};
  \node[draw=none,fill=none] (not-vn) at (0.8, -2*0.8) {};

  \foreach \i in {1,2,3}
    \node[cn] (cn2\i) at (2*0.8*\i + 0.8, -3*0.8) {};

  \draw (vn11) -- (cn11);
  \draw (vn12) -- (cn12);
  \draw (vn13) -- (cn11);
  \draw (vn13) -- (cn12);
  \draw (vn14) -- (cn13);
  \draw (vn15) -- (cn11);
  \draw (vn15) -- (cn13);
  \draw (vn16) -- (cn12);
  \draw (vn16) -- (cn13);
  \draw (vn17) -- (cn11);
  \draw (vn17) -- (cn12);
  \draw (vn17) -- (cn13);

  \draw (vn21) -- (cn21);
  \draw (vn22) -- (cn22);
  \draw (vn23) -- (cn21);
  \draw (vn23) -- (cn22);
  \draw (vn24) -- (cn23);
  \draw (vn25) -- (cn21);
  \draw (vn25) -- (cn23);
  \draw (vn26) -- (cn22);
  \draw (vn26) -- (cn23);
  \draw (vn27) -- (cn21);
  \draw (vn27) -- (cn22);
  \draw (vn27) -- (cn23);

  \draw (vn12) -- (mac2) -- (vn21);
  \draw (vn13) -- (mac3) -- (vn22);
  \draw (vn14) -- (mac4) -- (vn23);
  \draw (vn15) -- (mac5) -- (vn24);
  \draw (vn16) -- (mac6) -- (vn25);
  \draw (vn17) -- (mac7) -- (vn26);

  \draw[line width=.7, dashed, KITgreen] (vn11) -- ([yshift=-.3cm] not-vn.south);
  \draw[-latex, line width=.7, KITgreen] (not-vn.center) --node[below=.15cm,fill=none]{\textcolor{KITgreen}{$\tau$}} ([xshift=-.1cm] vn21.west);
\end{tikzpicture}
  \vspace*{-.1cm}
  \caption{Exemplary joint graph with a delay $\tau = 1$. \acp{VN} are represented by circles, \acp{CN} are squares and \acp{MN} are triangles.}
  \label{fig:joint-graph}
\end{figure}
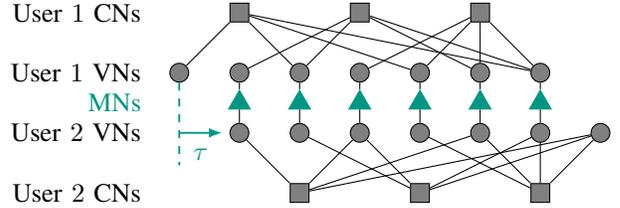

The following analogy to the \ac{BEC} allows the use of well-studied decoding algorithms~\cite{fengler_advantages_2023}:
a received $Y_i \in \{\pm 2\}$ defines both individual \ac{BPSK} symbols $X^{(1)}_i$ and $X^{(2)}_{i-\tau}$ unambiguously.
However, a received $Y_i = 0$ causes ambiguity between the two options ${X^{(1)}_i = -1}$, ${X^{(2)}_{i-\tau} = 1}$ and ${X^{(1)}_i = 1}$, ${X^{(2)}_{i-\tau} = -1}$.
For  $i \leq \tau$, a received ${Y_i = \pm 1}$ equals the transmitted symbol of the earlier transmitter, while for $i > n$, a received $Y_i = \pm 1$ equals the transmitted symbol of the delayed transmitter.
Using these observations, vectors containing known bits and erasures are constructed for both sent codewords.
Subsequently, these vectors are decoded by leveraging \ac{BEC} decoding techniques, like \emph{peeling decoding}~\cite{richardson_modern_2008}.
Additionally, decoding of both codewords is aided by $X^{(1)}_i \neq X^{(2)}_{i-\tau}$ if $Y_i = 0$.
Following~\cite{fengler_advantages_2023}, we model this knowledge using \acp{MN} that connect to one \ac{VN} in each peeling decoding.
If a \ac{VN} with $Y_i=0$ is recovered in the decoder of one user, its negated value is propagated to the \ac{VN} in the decoder of the other user via the \ac{MN} connecting the two \acp{VN}.
Fig.~\ref{fig:joint-graph} shows the factor graph corresponding to this setup consisting of two Tanner graphs interconnected by \acp{MN}.
We call this graph the \emph{joint graph}.
Even though the degree of \acp{VN} connected to \iac{MN} is incremented compared to the single-user Tanner graph, we only count the edges in the Tanner graph of a single user to define the degree of a \ac{VN}.
This maintains consistency with literature on \ac{LDPC} codes.

To evaluate decoding performance, we use the \ac{PUPE} which is introduced in~\cite{polyanskiy_perspective_2017} and is defined as
\begin{equation}
  \text{PUPE} = \frac{1}{2} P\left(\bm{X}^{(1)} \notin \mathcal{L}\right)
  + \frac{1}{2} P\left(\bm{X}^{(2)} \notin \mathcal{L}\right),
\end{equation}
where $\mathcal{L}$ is the set of codewords output by the decoding algorithm.
In this paper, $\mathcal{L}$ contains two codewords if decoding converges, else it is empty.
Note, that attributing messages to the transmitters is not required due to the unsourced paradigm.

Peeling decoding stops if an erased \emph{stopping set} is detected.
A stopping set is a set of \acp{VN}, such that any factor node (i.e., \ac{MN} or \ac{CN}) is connected to at least two \acp{VN} in the set~\cite{richardson_modern_2008}.
For the joint decoding approach, these stopping sets manifest in the joint graph.
In~\cite{fengler_advantages_2023}, the authors describe a family of stopping sets consisting of $4K$ degree-one \acp{VN}, with $K\in \mathbb{N}$.
In particular, small stopping sets with $K=1$ form with a high probability and induce a high \ac{PUPE}.
The next section analyzes the emergence and removal of these stopping sets.

\section{Removal of Small Weight Stopping Sets} \label{sec:analysis-removal}

We start the analysis of stopping sets by noting a difference between single-user decoding and asynchronous two-user decoding: the order of the code bits, i.e., the order of the columns in the \ac{PCM} $\bm{H}$, affects the decoding performance in the two-user case, but not in the single-user case.
For the discussion in this paper, we define the location of a \ac{VN} as the position of the bit corresponding to that \ac{VN} in the codeword.
The order of \acp{VN} and distances between \acp{VN} are defined likewise.
Each \ac{MN} connects a \ac{VN} in the graph of user $1$ with a \ac{VN} that is located $\tau$ code positions before it in the graph of user $2$.
Thus, changing the order of the \acp{VN} alters the graph structure and, therefore, the decoding performance.

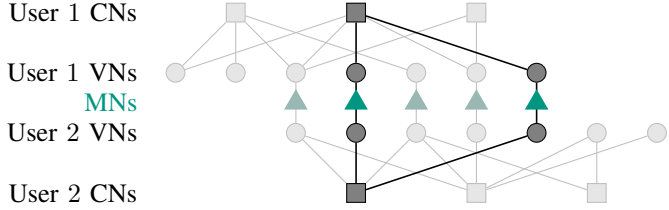
\begin{figure}[tbp]
  \centering
  \begin{tikzpicture}[%
  every node/.style={draw=lightgray, fill=gray!20!white, minimum size = 2.5mm, inner sep = 0},
  every path/.style={draw=lightgray},
  vn/.style={circle},
  cn/.style={rectangle},
  mac/.style={
    draw,
    color=KITgreen!20!lightgray,
    isosceles triangle,
    rotate = 90,
    isosceles triangle apex angle = 60,
  },
  macemph/.style={
    draw,
    color=KITgreen,
    isosceles triangle,
    rotate = 90,
    isosceles triangle apex angle = 60,
  },
  label/.style={draw=none, fill=none, anchor = east},
  emph/.style={draw=black, fill=gray},
  lineemph/.style={draw=black, line width=0.6},
]

  \node[label] at (.3,0*0.8) {User $1$ CNs};
  \node[label] at (.3,-1*0.8) {User $1$ VNs};
  \node[label, color=KITgreen] at (.3,-1.5*0.8) {MNs};
  \node[label] at (.3,-2*0.8) {User $2$ VNs};
  \node[label] at (.3,-3*0.8) {User $2$ CNs};

  \node[cn] (cn11) at (2*0.8,0) {};
  \node[cn, emph] (cn12) at (4*0.8,0) {};
  \node[cn] (cn13) at (6*0.8,0) {};

  \node[vn] (vn11) at (1*0.8, -1*0.8) {};
  \node[vn] (vn12) at (2*0.8, -1*0.8) {};
  \node[vn] (vn13) at (3*0.8, -1*0.8) {};
  \node[vn, emph] (vn14) at (4*0.8, -1*0.8) {};
  \node[vn] (vn15) at (5*0.8, -1*0.8) {};
  \node[vn] (vn16) at (6*0.8, -1*0.8) {};
  \node[vn, emph] (vn17) at (7*0.8, -1*0.8) {};

  \node[vn] (vn21) at (3*0.8, -2*0.8) {};
  \node[vn, emph] (vn22) at (4*0.8, -2*0.8) {};
  \node[vn] (vn23) at (5*0.8, -2*0.8) {};
  \node[vn] (vn24) at (6*0.8, -2*0.8) {};
  \node[vn, emph] (vn25) at (7*0.8, -2*0.8) {};
  \node[vn] (vn26) at (8*0.8, -2*0.8) {};
  \node[vn] (vn27) at (9*0.8, -2*0.8) {};

  \node[cn, emph] (cn21) at (4*0.8, -3*0.8) {};
  \node[cn] (cn22) at (6*0.8, -3*0.8) {};
  \node[cn] (cn23) at (8*0.8, -3*0.8) {};

  \draw (vn13) -- (vn21);
  \draw[lineemph] (vn14) -- (vn22);
  \draw (vn15) -- (vn23);
  \draw (vn16) -- (vn24);
  \draw[lineemph] (vn17) -- (vn25);

  \node[mac] (mac3) at (3*0.8, -1.5*0.8) {};
  \node[macemph] (mac4) at (4*0.8, -1.5*0.8) {};
  \node[mac] (mac5) at (5*0.8, -1.5*0.8) {};
  \node[mac] (mac6) at (6*0.8, -1.5*0.8) {};
  \node[macemph] (mac7) at (7*0.8, -1.5*0.8) {};

  \draw (vn11) -- (cn11);
  \draw (vn11) -- (cn12);
  \draw (vn12) -- (cn11);
  \draw (vn13) -- (cn11);
  \draw (vn13) -- (cn12);
  \draw (vn13) -- (cn13);
  \draw (vn15) -- (cn11);
  \draw (vn16) -- (cn12);
  \draw (vn16) -- (cn13);
  \draw[lineemph] (vn14) -- (cn12);
  \draw[lineemph] (vn17) -- (cn12);

  \draw (vn21) -- (cn21);
  \draw (vn21) -- (cn22);
  \draw (vn23) -- (cn21);
  \draw (vn23) -- (cn22);
  \draw (vn23) -- (cn23);
  \draw (vn24) -- (cn22);
  \draw (vn26) -- (cn22);
  \draw (vn26) -- (cn23);
  \draw (vn27) -- (cn22);
  \draw[lineemph] (vn22) -- (cn21);
  \draw[lineemph] (vn25) -- (cn21);

\end{tikzpicture}
  \vspace*{-.3cm}
  \caption{A joint graph for $\tau = 2$ with a highlighted 4SET.}
  \label{fig:4set}
\end{figure}

Fig.~\ref{fig:4set} depicts the smallest possible stopping set in the joint graph, which we name the \acf{4SET}.
A 4SET consists of two pairs of degree-one \acp{VN}, one pair from each of the interconnected Tanner graphs.
Hereby, both \acp{VN} in a pair are connected to the same \ac{CN}.
Each \ac{VN} is connected to a \ac{VN} in the other pair via \iac{MN}.

Due to its small number of edges, the \ac{4SET} has the highest multiplicity and, hence, it is the most probable stopping set to form in the joint graph of randomly generated codes.
Furthermore, if each code bit takes the values $0$ or $1$ with equal probability, the \ac{4SET} is erased with probability $\frac{1}{4}$, inducing a high \ac{PUPE}~\cite{fengler_advantages_2023}.
In the following, we analyze the conditions in which \acp{4SET} form and introduce a simple method of eliminating them for all delays $\tau$.
Subsequently, we discuss cases in which this simple method is insufficient and demonstrate an approach capable of eliminating \acp{4SET} for a significantly broader range of code ensembles.

\subsection{Properties of \acp{4SET}}

To describe the formation of \acp{4SET}, we use multisets denoted by $\{\cdot\}_\mathrm{m}$.
The multiplicity function $m_\mathcal{X}(x)$ counts how often the multiset $\mathcal{X}$ contains $x$ and the operand $\uplus$ denotes a union of multisets where multiplicities of the individual multisets are summed, e.g., $\{a, b\}_\mathrm{m} \uplus \{a, a, b\}_\mathrm{m} = \{a, a, a, b, b\}_\mathrm{m}$.

We define the set $\mathcal{V}_j \subset \{1,\dots,n\}$ to consist of the locations of degree-one \acp{VN} connected to \ac{CN}~$c_j$.
Furthermore, we define the distance multisets $\mathcal{D}_j$ per \ac{CN}~$c_j$ and the total distance multiset $\mathcal{D}$:
\begin{align}
  \mathcal{D}_j &= \big\{|k - \ell|: k,\ell \in \mathcal{V}_j, k > \ell\big\}_\mathrm{m} \quad \text{and} \\
  \mathcal{D} &= \biguplus_{j=1}^m \mathcal{D}_j.
\end{align}

\emph{Example:} For the two Tanner graphs in the joint graph in Fig.~\ref{fig:4set}, the degree-one \ac{VN} sets per \ac{CN} are $\mathcal{V}_1 = \{2,5\}$, ${\mathcal{V}_2 = \{4,7\}}$ and $\mathcal{V}_3 = \emptyset$.
As a result the distance multisets are $\mathcal{D}_1 = \{3\}_\mathrm{m}$, $\mathcal{D}_2 = \{3\}_\mathrm{m}$ and $\mathcal{D}_3 = \{\}_\mathrm{m}$, leading to ${\mathcal{D} = \{3,3\}_\mathrm{m}}$.
Both degree-one \ac{VN} pairs $\mathcal{V}_1$ and $\mathcal{V}_2$ lead to the same distance~$3$.
Consequently, there exists a delay $\tau$ such that the two \ac{VN} pairs align to form the \ac{4SET} shown in Fig.~\ref{fig:4set}.

We formalize this notion in the following theorems.
\begin{theorem}[\ac{4SET}-free \ac{PCM}]
  Given a \ac{PCM} $\bm{H}$, if the corresponding total distance multiset $\mathcal{D}$ contains only unique elements, no \acp{4SET} can form in the joint graph for any delay $\tau > 0$.
  We call this a \ac{4SET}-free \ac{PCM}.
  \label{theorem:4set-free-pcm}
\end{theorem}
\begin{IEEEproof}
  A \ac{4SET} contains two pairs of degree-one \acp{VN} $\{i_1, i_1 + d\}$ and $\{i_2, i_2 + d\}$, such that both \acp{VN} in the respective pair connect to the same \ac{CN}.
  If $\tau \geq 1$, these must be two distinct pairs, i.e., $i_1 \neq i_2$ and at least one entry in $\mathcal{D}$ is not unique as $m_\mathcal{D}(d) \geq 2$.
  Therefore, if all entries of $\mathcal{D}$ are unique, there are no \acp{4SET}.
\end{IEEEproof}
\begin{theorem}[Converse]
  If the total difference multiset $\mathcal{D}$ contains non-unique elements, there exists a delay $\tau \geq 1$, such that a \ac{4SET} forms in the joint graph.
\end{theorem}
\begin{IEEEproof}
  Let $d$ be the non-unique element of $\mathcal{D}$.
  There exist at least two distinct pairs $\{i_1, i_1 + d\}$ and $\{i_2, i_2 + d\}$ of degree-one \acp{VN}, such that both \acp{VN} in a pair connect to the same \ac{CN}.
  Without loss of generality, we assume $i_1 > i_2$ and find that a \ac{4SET} forms for $\tau = i_1 - i_2$.
\end{IEEEproof}

According to Theorem~\ref{theorem:4set-free-pcm}, the search for a \ac{4SET}-free \ac{PCM} can be expressed as the search for a total distance multiset $\mathcal{D}$ with unique elements.
Ensuring that the elements of $\mathcal{D}$ are unique is simplified if $|\mathcal{D}|$ is reduced.
\begin{theorem}
  For a fixed number of degree-one \acp{VN} and a fixed number of \acp{CN}, $|\mathcal{D}|$ is minimized if, ${\big||\mathcal{V}_u| - |\mathcal{V}_w|\big| \leq 1}$, $\forall u,w \in [m]$.
  We refer to this as an even distribution of degree-one \acp{VN} across all \acp{CN}.
  \label{theorem:minimize-d}
\end{theorem}
\begin{IEEEproof}
  For the proof see~%
  Appendix~\ref{apx:proof-min-d}. %
\end{IEEEproof}

\begin{figure}[tbp]
  \centering
  \begin{tikzpicture}
  \tikzset{
    common/.style = {
      line width=1.1pt,
      color = KITblue,
      mark=none
    },
    v1/.style = {
      color = KITred,
    },
  }
  \begin{axis}[
      width=3cm,
      height=3cm,
      axis x line = bottom,
      axis y line = left,
      axis line style = { -latex },
      scale only axis,
      ymin = 0,
      ymax = 1.2,
      xmin = 0, 
      xmax = 1.2, 
      xlabel={$L_1$},
      yminorticks=true,
      ylabel={$r_\mathrm{d}$},
      every axis x label/.style={
        at = {(ticklabel* cs:1)},
        anchor = north west
      },
      every axis y label/.style={
        at = {(ticklabel* cs:1)},
        anchor = west
      },
      legend style={
        at={(1.1,0.5)},
        anchor=west,
        legend cell align=left,
        align=left,
        draw=none,
        row sep=5,
        column sep=3,
        font=\footnotesize        
      }
    ]%

    \addplot[common, black, dashed, name path=capacity, domain=0:1] {0.75};%
    \addlegendentry{AU-BAC capacity}

    \addplot[common, v1, name path=v1, domain=0:1] {1 - x / 1};%
    \addlegendentry{$V = 1$}

    \addplot[common, name path=v2, domain=0:1] {1 - x / 2};%
    \addlegendentry{$V = 2$}

    \path[name path=axis] (0,0) -- (100,0);

    \addplot[KITblue, only marks, mark=x, draw=none,every mark/.append style={scale=1.5,line width=1pt}] coordinates {(0.297,0.65)};
    \addlegendentry{Ensemble~\ref{ensemble-1}}

    \addplot[KITblue, only marks, mark=asterisk, draw=none,every mark/.append style={scale=1.5,line width=1pt}] coordinates {(0.41,0.667)};
    \addlegendentry{Ensemble~\ref{ensemble-2}}

    \addplot[KITblue, only marks, mark=10-pointed star, draw=none,every mark/.append style={scale=1.5,line width=1pt}] coordinates {(0.746,0.708)};
    \addlegendentry{Ensemble~\ref{ensemble-3}}

    \path[name path=bound-v1, intersection segments={of=v1 and capacity, sequence=R1 -- L2}];
    \addplot[pattern={Lines[angle=110]}, pattern color=white!30!KITred] fill between [of= axis and bound-v1, soft clip={domain=0:1}];

    \path[name path=bound-v2, intersection segments={of=v2 and capacity, sequence=R1 -- L2}];
    \addplot[pattern={Lines[angle=20]}, pattern color=white!30!KITblue] fill between [of= axis and bound-v2, soft clip={domain=0:1}];

  \end{axis}%
\end{tikzpicture}\\
  \vspace*{-.3cm}
  \caption{$(L_1,r_\mathrm{d})$-region with different ensembles and values for $V$ marked.}
  \label{fig:l1-r-even-dist}
\end{figure}
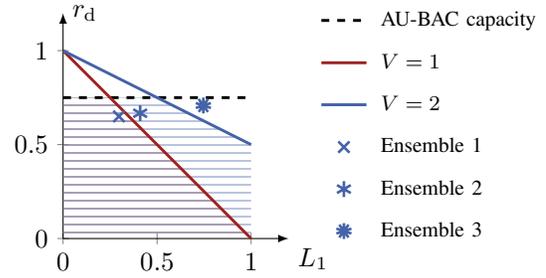

We define the average number of degree-one \acp{VN} connected to each \ac{CN} as
\begin{equation}
  V \coloneq \frac{L_1}{1 - r_\mathrm{d}}.
  \label{eq:avg-deg-one-vn-per-cn}
\end{equation}
If $V\leq1$, evenly distributing degree-one \acp{VN} across \acp{CN} ensures $|\mathcal{V}_j| \leq 1$ for all \acp{CN} $c_j$, i.e., $|\mathcal{D}| = 0$, resulting in a \ac{4SET}-free \ac{PCM}.
Fig.~\ref{fig:l1-r-even-dist} highlights the {$(L_1,r_\mathrm{d})$\nobreakdash-region} in which $V \leq 1$ and this approach is applicable.

\begin{table}%
\centering
  \caption{Degree Distributions Obtained Using the Optimization from~\cite{fengler_advantages_2023}}
  \label{tab:degree-distributions}
  \vspace*{-.2cm}
  \begin{tabular}{r S S S}
    \toprule
     & {Ensemble~\enslabel{ensemble-1}} & {Ensemble~\enslabel{ensemble-2}} & {Ensemble~\enslabel{ensemble-3}} \\
    \midrule
    $V$ & 0.85 & 1.2 & 2.55 \\
    $r_\mathrm{d}$ & 0.65 & 0.667 & 0.708 \\
    \midrule
    $L_1$ & 0.297 & 0.41 & 0.746 \\
    $L_2$ & 0.703 & 0.59 & 0.254 \\
    $R_3$ & 0.064 & {---} & {---} \\
    $R_4$ & 0.789 & 0.87 & 0.706 \\
    $R_5$ & {---} & {---}  & 0.294 \\
    $R_{10}$ & 0.147 & 0.13 & {---} \\
    \bottomrule
  \end{tabular}
\end{table}
Table~\ref{tab:degree-distributions} presents a set of degree distributions optimized using the methods proposed in~\cite{fengler_advantages_2023} and chosen to demonstrate a wide range of values for~$V$.
Each pair of \ac{VN} and \ac{CN} degree distributions occupies a point in the $(L_1,r_\mathrm{d})$-plane.
The points for the three degree distributions in Table~\ref{tab:degree-distributions} are shown in Fig.~\ref{fig:l1-r-even-dist}.
Note that the approaches proposed in this paper add more structure to the constructed codes, e.g., by distributing the degree-one \acp{VN} evenly across all \acp{CN}.
Thus, these codes form a subset of the full ensemble of random edge permutations for a given pair of degree distributions.
The \ac{DE} in~\cite{fengler_advantages_2023} is derived for codes from this unconstrained ensemble and is, therefore, not valid for our codes.
However, we expect the differences to be minimal and our simulation results show that our proposed methods achieve low error probabilities with degree distributions optimized following the \ac{DE}.
In the next subsection, we propose how to deal with $V>1$.

\subsection{Refined Removal of \acp{4SET}}

A \ac{4SET}-free \ac{PCM} can be created for Ensemble~\ref{ensemble-1} by distributing degree-one \acp{VN} across \acp{CN} evenly, as it has $V \approx 0.89 \leq 1$.
However, more refined methods must be used for the two remaining degree distribution pairs.

For $V\leq 2$ it is possible to give a deterministic construction that fully avoids \acp{4SET}.
We can construct a \ac{4SET}-free \ac{PCM} for an ensemble with $L_1 = 1$ and $r_\mathrm{d}=\frac{1}{2}$ by choosing $\mathcal{V}_j = \left\{ \frac{n+1}{2} \pm \left(j - \frac{1}{2}\right)\right\}$.
This construction can be extended to find \ac{4SET}-free \acp{PCM} for ensembles with $V\leq 2$ (see~%
Appendix~\ref{apx:v2pcm}) %
and allows us to find a \ac{4SET}-free \ac{PCM} for Ensemble~\ref{ensemble-2}.
However, for $V >2$ no such deterministic solution is known to the authors.

\begin{algorithm}[tb]
  \caption{Convert a \ac{PCM} $\bm{H}$ into a \ac{4SET}-free \ac{PCM}}
  \label{alg:4set-removal-alg}
  \begin{algorithmic}
    \REQUIRE Parity check matrix $\bm{H} \in \mathbb{F}_2^{m \times n}$
    \STATE $\mathcal{D}' \coloneq \{\}_\text{m}$
    \FOR{$j = 1$ to $m$}
      \STATE $\mathcal{V}_j' \coloneq \emptyset$
      \WHILE{$|\mathcal{V}_j'| <$ \# degree-one VNs connected to CN~$c_j$}
        \STATE Choose random $k\in[n] \setminus \bigcup_{j' \leq j} \mathcal{V}_{j'}'$
        \vspace*{0.25em}
        \IF{elements of $\{|k - \ell|: \ell \in \mathcal{V}_j'\}_\text{m} \uplus \mathcal{D}'$ unique}
          \STATE $\mathcal{V}_j' \coloneq \mathcal{V}_j' \cup \{k\}$
          \STATE $\mathcal{D}' \coloneq \mathcal{D}' \uplus \{|k-\ell|: \ell \in \mathcal{V}_j'\}_\text{m}$
        \ENDIF
        \IF{max. number of iterations reached}
          \STATE Failed to convert $\bm{H}$ into a \ac{4SET}-free \ac{PCM}
        \ENDIF
      \ENDWHILE
    \ENDFOR
    \STATE Permute columns of $\bm{H}$ according to $\{\mathcal{V}_1',\dots, \mathcal{V}_m'\}$
    \RETURN $\bm{H}$
  \end{algorithmic}
\end{algorithm}

Instead, we propose the stochastic Algorithm~\ref{alg:4set-removal-alg} to optimize \acp{PCM} sampled from degree-distributions such as Ensemble~\ref{ensemble-3}.
Consider a \ac{PCM} $\bm{H} \in \mathbb{F}_2^{m\times n}$.
Then, for each \ac{CN}~$c_j$, this algorithm iteratively samples degree-one \ac{VN} locations $k$ and checks whether adding $k$ to the new degree-one \ac{VN} locations per \ac{CN} set $\mathcal{V}_j'$ would create duplicates in the new total distance multiset $\mathcal{D}'$.
Location $k$ is only added to $\mathcal{V}_j'$ if it causes no duplicates.
After all $\mathcal{V}_1',\dots,\mathcal{V}_m'$ are constructed, the columns of the \ac{PCM} $\bm{H}$ are permuted such that the \ac{VN} locations in the new \ac{PCM} match the \ac{VN} locations in the found $\mathcal{V}_1',\dots,\mathcal{V}_m'$.
As the total distance multiset of the resulting \ac{PCM} only contains unique elements, the returned \ac{PCM} is \ac{4SET}-free, i.e., no \ac{4SET} can form in the joint graph for any $\tau > 0$.

Our simulations show that this algorithm produces \ac{4SET}-free \acp{PCM} in most cases of interest, even for initial \acp{PCM} with degree-one \acp{VN} not distributed evenly across all \acp{CN}.
One advantage of the proposed algorithm is that it works by only permuting the columns of the \ac{PCM}.
Thus, all single-user properties of the \ac{PCM} remain unaffected.

To analyze the effectiveness of the proposed algorithm, we bound the maximum possible $(L_1,r_\mathrm{d})$-region for which \ac{4SET}-free \acp{PCM} exists.
The size of the total distance multiset is lower bounded by (see~%
Appendix~\ref{apx:lagrange}) %
\begin{equation}
  |\mathcal{D}| \geq m \cdot \frac{V (V - 1)}{2}.
  \label{eq:bound-on-cardinality-d}
\end{equation}
Since distances larger than $n-1$ cannot occur, the elements of $\mathcal{D}$ can only take on values in ${[n-1]}$.
Hence, if $\mathcal{D}$ only contains unique elements (see Theorem~\ref{theorem:4set-free-pcm}),

\begin{equation}
  n - 1 \geq |\mathcal{D}| \geq m \cdot \frac{V (V - 1)}{2}.
\end{equation}
Using~\eqref{eq:avg-deg-one-vn-per-cn} and $m = n(1-r_\mathrm{d})$ yields the asymptotic upper bound
\begin{equation}
   r_\mathrm{d} \leq 1 - \frac{L_1^2}{2 + L_1}.
\end{equation}

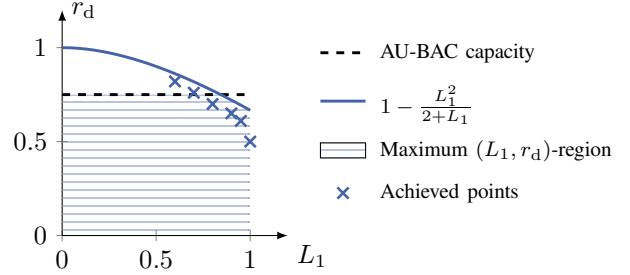
\begin{figure}[tbp]
  \centering
  \begin{tikzpicture}
  \tikzset{
    common/.style = {
      line width=1.1pt,
      color = KITblue,
      mark=none
    },
    v1/.style = {
      color = KITred,
    },
  }
  \begin{axis}[
      width=3cm,
      height=3cm,
      axis x line = bottom,
      axis y line = left,
      axis line style = { -latex },
      scale only axis,
      ymin = 0,
      ymax = 1.2,
      xmin = 0, 
      xmax = 1.2, 
      xlabel={$L_1$},
      yminorticks=true,
      ylabel={$r_\mathrm{d}$},
      every axis x label/.style={
        at = {(ticklabel* cs:1)},
        anchor = north west
      },
      every axis y label/.style={
        at = {(ticklabel* cs:1)},
        anchor = west
      },
      legend style={
        at={(1.1,0.5)},
        anchor=west,
        legend cell align=left,
        align=left,
        draw=none,
        row sep=5,
        column sep=3,
        font=\footnotesize
      }
    ]%

    \addplot[common, black, dashed, name path=capacity, domain=0:1] {0.75};%
    \addlegendentry{AU-BAC capacity}

    \addplot[common, name path=curve, domain=0:1] {1 - x^2/(2+x)};%
    \addlegendentry{$1 - \frac{L_1^2}{2 + L_1}$}

    \path[name path=axis] (0,0) -- (100,0);

    \path[name path=bound, intersection segments={of=curve and capacity, sequence=R1 -- L2}];
    \addplot[pattern={Lines[angle=20]}, pattern color=white!30!KITblue] fill between [of= axis and bound, soft clip={domain=0:1}];
    \addlegendentry{Maximum $(L_1, r_\mathrm{d})$-region}

    \addplot[KITblue, only marks, mark=x, draw=none,every mark/.append style={scale=1.5,line width=1pt}] coordinates {(0.6,0.82) (0.7,0.76) (0.8,0.7) (0.9,0.65) (0.95,0.61) (1.0,0.5)};
    \addlegendentry{Achieved points}
  \end{axis}%
\end{tikzpicture}\\
  \vspace*{-.3cm}
  \caption{$(L_1,r_\mathrm{d})$-region with bound on points for which a \ac{4SET} free \ac{PCM} exists.}
  \label{fig:l1-r-bound}
\end{figure}

Fig.~\ref{fig:l1-r-bound} illustrates this bound in the $(L_1,r_\mathrm{d})$-region.
No \ac{4SET}\nobreakdash-free \ac{PCM} can be found for ensembles outside the bound.
Yet, it remains an open question whether the bound is achievable.
Additionally, Fig.~\ref{fig:l1-r-bound} contains points which we achieved using Algorithm~\ref{alg:4set-removal-alg}.
A small gap between the achieved points and the upper bound remains and could possibly be reduced by improving Algorithm~\ref{alg:4set-removal-alg}.
However, Fig.~\ref{fig:l1-r-bound} demonstrates that, heuristically, Algorithm~\ref{alg:4set-removal-alg} is capable of finding \ac{4SET}-free \acp{PCM} for the vast majority of $(L_1,r_\mathrm{d})$\nobreakdash-pairs.

\begin{algorithm}[tb]
  \caption{Degree-one \ac{VN} stopping set removal}
  \label{alg:deg-one-vn-ss-removal-alg}
  \begin{algorithmic}
    \REQUIRE Parity check matrix $\bm{H} \in \mathbb{F}_2^{m \times n}$
    \REPEAT
      \STATE Find set of $\mathcal{V}_j$ using Algorithm~\ref{alg:4set-removal-alg}
    \UNTIL{All $\mathcal{V}_j$ cause no larger degree-one VN stopping-sets}
    \STATE Permute columns of $\bm{H}$ according to $\{\mathcal{V}_1,\dots, \mathcal{V}_m\}$
    \RETURN $\bm{H}$
  \end{algorithmic}
\end{algorithm}

Finally, larger degree-one \ac{VN} stopping sets have a smaller multiplicity than the \ac{4SET} and are, hence, less probable to form in the joint graph.
A \ac{PCM} with no degree-one \ac{VN} stopping sets can typically be found by running Algorithm~\ref{alg:4set-removal-alg} sufficiently often.
For instance, this is possible for Ensembles~\ref{ensemble-1} and~\ref{ensemble-2}, with $10$ to $20$ runs of Algorithm~\ref{alg:4set-removal-alg}. For Ensemble~\ref{ensemble-3}, this approach is unsuccessful.
One way of testing a \ac{PCM} for degree-one \ac{VN} stopping sets is explained in~%
Appendix~\ref{apx:finding-deg-one-vn-ss}. %
Algorithm~\ref{alg:deg-one-vn-ss-removal-alg} summarizes the iterative application of Algorithm~\ref{alg:4set-removal-alg} to remove degree-one \ac{VN} stopping sets of any size.

\section{Numerical Results} \label{sec:results}

Theorem~\ref{theorem:minimize-d} states that degree-one \acp{VN} which are distributed evenly across all \acp{CN} minimize the total distance multiset size.
This reduces the possible number of non-unique distances, thereby making the removal of \acp{4SET} easier.
Thus, to refine the \acf{RCC} used in~\cite{fengler_advantages_2023}, for which all edges in the Tanner graph are randomly permuted, we propose \emph{Even-\ac{RCC}}, in which degree-one \acp{VN} are first distributed randomly, but evenly, across all \acp{CN}.
Thereafter, the remaining edges are added randomly.
An alternative to \ac{RCC} is \acf{PEG}, which constructs codes while greedily maximizing the local girth when placing each edge~\cite{hu_regular_2005}.
In this way, it reduces the number of short cycles and improves decoding performance in the single-user case.
We use the \ac{PEG} implementation published on~\cite{mackay_peg_2008}.
As \ac{PEG} first processes \acp{VN} of low degrees and chooses the \ac{CN} with the lowest degree in these first iterations, it inherently distributes degree-one \acp{VN} evenly across all \acp{CN}.
This, in combination with its known single-user advantages, makes \ac{PEG} a promising code construction technique for the \ac{AU-BAC}.
Since Algorithm~\ref{alg:deg-one-vn-ss-removal-alg} only permutes columns, it does not alter the structure of the input \ac{PCM}.
Thus Even-\ac{RCC} as well as \ac{PEG} can both be combined with Algorithm~\ref{alg:deg-one-vn-ss-removal-alg} as a second step.

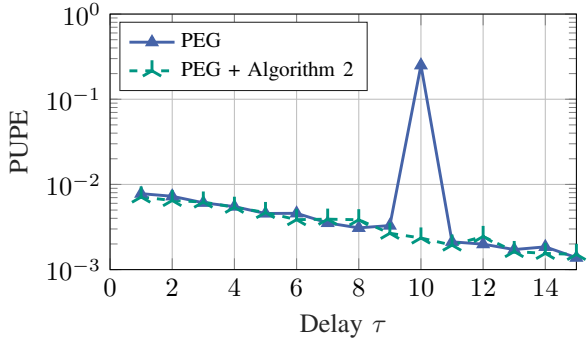
\begin{figure}[tbp]
  \centering
  \begin{tikzpicture}
  \tikzset{
    common/.style = {line width=1.1pt},
    random/.style = {KITred, solid, mark=square*, mark options={solid}},
    even random/.style = {KITorange, solid, mark=diamond*, mark options={solid}},
    peg/.style = {KITblue, solid, mark=triangle*, mark options={solid}},
    peg_alg/.style = {KITgreen, solid, mark=Mercedes star, mark options={solid}, mark size=.4em},
    algorithm/.style = {dashed, mark options={fill=white,solid}},
    no mark/.style = {mark=none},
    hide/.style = {draw=none,mark=none},
  }
  \begin{axis}[
     width=.7\columnwidth,
     height=3.4cm,
     at={(0.758in,0.645in)},
     scale only axis,
     xlabel style={font=\color{white!15!black}},
     xlabel={Delay $\tau$},
     ymode=log,
     ymin=1e-3,
     ymax=1,
     xmin=0,
     xmax=15,
     yminorticks=true,
     ylabel style={font=\color{white!15!black}},
     ylabel={PUPE},
     axis background/.style={fill=white},
     xmajorgrids,
     ymajorgrids,
     legend style={
       at={(0.02,0.97)},
       anchor=north west,
       legend cell align=left,
       align=left,
       draw=white!15!black,
       font=\footnotesize
     }
   ]%

   \addplot+[%
     common,
     peg,
   ]table[
     x = tau,
     y = with-ss-0,
     col sep = comma
   ]{data/pupe_vs_tau_n500_r0.667.csv};%
   \addlegendentry{\ac{PEG}}

   \addplot+[%
     common,
     peg_alg,
     algorithm,
   ]table[
     x = tau,
     y = no-ss-0,
     col sep = comma
   ]{data/pupe_vs_tau_n500_r0.667.csv};%
   \addlegendentry{\ac{PEG} + Algorithm~\ref{alg:deg-one-vn-ss-removal-alg}}

  \end{axis}%
\end{tikzpicture}%
  \vspace*{-.3cm}
  \caption{\ac{PUPE} of a $n=500$ code constructed from Ensemble~\ref{ensemble-2} using \ac{PEG} for different values of $\tau$. \num{100000} transmissions are simulated per point. The \ac{PUPE} peak at $\tau = 10$ is caused by a \ac{4SET}.}
  \label{fig:pupe-vs-tau}
\end{figure}

Fig.~\ref{fig:pupe-vs-tau} demonstrates the effect of a \ac{4SET} by tracking the \ac{PUPE} of one code realization from Ensemble~\ref{ensemble-2} for different delays $\tau$.
For $\tau = 10$, a \ac{4SET} forms in the joint graph  and the \ac{PUPE} spikes to $\frac{1}{4}$.
Such spikes increase the average \ac{PUPE} in a system with random delays.
We employ Algorithm~\ref{alg:deg-one-vn-ss-removal-alg} to successfully remove the \ac{4SET}.
Fig.~\ref{fig:pupe-vs-tau} shows that this mitigates the spike, significantly reducing the average \ac{PUPE}.

\begin{figure}[tbp]
  \centering
  \begin{tikzpicture}
  \tikzset{
    common/.style = {line width=1.1pt},
    random/.style = {KITorange, solid, mark=square*, mark options={solid}},
    even random/.style = {KITred, solid, mark=diamond*, mark options={solid}},
    even random_alg/.style = {KITpurple, solid, mark=+, mark options={solid}, mark size=.4em},
    peg/.style = {KITblue, solid, mark=triangle*, mark options={solid}},
    peg_alg/.style = {KITgreen, solid, mark=Mercedes star, mark options={solid}, mark size=.4em},
    algorithm/.style = {dashed, mark options={fill=white,solid}},
    no mark/.style = {mark=none},
    hide/.style = {draw=none,mark=none},
  }
  \begin{axis}[
     width=.7\columnwidth,
     height=4cm,
     at={(0.758in,0.645in)},
     scale only axis,
     xlabel style={font=\color{white!15!black}},
     xlabel={Block length $n$},
     ymode=log,
     xmode=log,
     ymin=1e-5,
     ymax=1e-1,
     xmin=100,
     xmax=1600,
     xtick={100, 200, 400, 800, 1600},
     xticklabels={$100$, $200$, $400$, $800$, $1600$},
     yminorticks=true,
     ylabel style={font=\color{white!15!black}},
     ylabel={PUPE},
     axis background/.style={fill=white},
     xmajorgrids,
     ymajorgrids,
     legend style={
       at={(0.02,0.03)},
       anchor=south west,
       legend cell align=left,
       align=left,
       draw=white!15!black,
       font=\footnotesize
     }
   ]%

   \addplot+[%
     common,
     random,
   ]table[
     x = n,
     y = pupe-rcc,
     col sep = comma
   ]{data/pupe_vs_n_short1_num_codes_2000.csv};%
   \addlegendentry{\ac{RCC}}

   \addplot+[%
     common,
     even random,
   ]table[
     x = n,
     y = pupe-rcc-even,
     col sep = comma
   ]{data/pupe_vs_n_short1_num_codes_2000_even_rcc.csv};%
   \addlegendentry{Even-\ac{RCC}}
   
   \addplot+[%
     common,
     peg,
   ]table[
     x = n,
     y = pupe-peg,
     col sep = comma
   ]{data/pupe_vs_n_short1_num_codes_2000.csv};%
   \addlegendentry{\ac{PEG}}

   \addplot+[%
     common,
     even random_alg,
     algorithm,
   ]table[
     x = n,
     y = pupe-rcc-even-alg2,
     col sep = comma
   ]{data/pupe_vs_n_short1_num_codes_2000_even_rcc.csv};%
   \addlegendentry{Even-\ac{RCC} + Algorithm~\ref{alg:deg-one-vn-ss-removal-alg}\hspace*{-.4em}}

   \addplot+[%
     common,
     peg_alg,
     algorithm,
   ]table[
     x = n,
     y = pupe-peg-alg2,
     col sep = comma
   ]{data/pupe_vs_n_short1_num_codes_2000.csv};%
   \addlegendentry{\ac{PEG} + Algorithm~\ref{alg:deg-one-vn-ss-removal-alg}}

  \end{axis}%
\end{tikzpicture}%
  \vspace*{-.3cm}
  \caption{Average \ac{PUPE} of $2000$ codes from Ensemble~\ref{ensemble-2} ($V\approx 1.2$) with a random delay $\tau\in[50]$ and for different block lengths $n$. $2000$ transmissions are simulated per code and block length.}
  \label{fig:pupe-vs-n}
\end{figure}
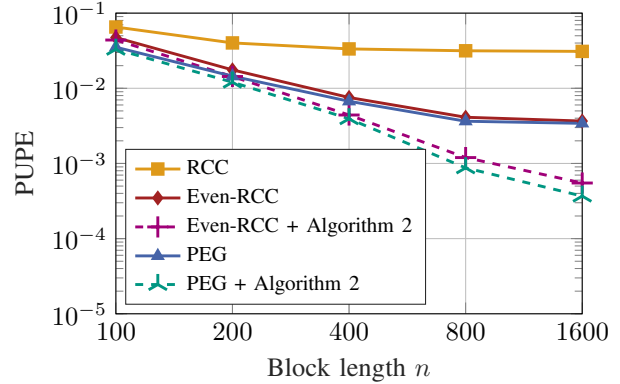

Fig.~\ref{fig:pupe-vs-n} shows the average \ac{PUPE} at different block lengths $n$, using a variety of construction methods.
For each simulated transmission, a random delay ${\tau\in[50]}$ is chosen and an average over $2000$ codes is computed.
Plain \ac{RCC} leads to an error floor at a \ac{PUPE} of $0.03$.
Many stopping sets occur with this construction method and the \ac{PUPE} does not drop with increasing block lengths.
In line with Theorem~\ref{theorem:minimize-d} and the expectation that minimal $|\mathcal{D}|$ typically leads to less \acp{4SET}, Even-\ac{RCC} reduces the error floor to a \ac{PUPE} of $0.0035$, i.e., by one order of magnitude.
Using \ac{PEG} instead of Even-\ac{RCC} yields only a small decrease in the \ac{PUPE}.
This hints that, for the \ac{AU-BAC}, the single-user properties of the used \ac{PCM} are significantly less important than multi-user properties such as the structure of the joint graph.
Finally, removing all degree-one \ac{VN} stopping sets leads to a further reduction in \ac{PUPE} and no visible error floor for the simulated block lengths.
At the highest simulated block length of $n=1600$, using \ac{PEG} with degree-one \ac{VN} stopping set removal reduces the \ac{PUPE} by a factor of $84.2$ compared to \ac{RCC} and a factor of $9.29$ compared to plain \ac{PEG}.
For the simulated ensemble and block lengths, transmission at a target \ac{PUPE} of $10^{-3}$ or below is only possible if degree-one \ac{VN} stopping sets are removed.

\section{Conclusion} \label{sec:conclusion}
In this work, we analyze the formation of small weight stopping sets in the joint graph of an \ac{AU-BAC}, which is a noise-free \ac{UMAC} with frame asynchrony.
We observe that the order of columns in the \ac{PCM} influences the decoding performance in the asynchronous, unsourced two-user case, while it is irrelevant in the single-user case.
Subsequently, we analyze the conditions in which a \ac{4SET} can occur in terms of \ac{VN} locations and distances between \acp{VN}.
We propose an algorithm to remove \acp{4SET} from a \ac{PCM} for all delays~$\tau \geq 1$ by simply permuting its columns.
Additionally, we note that distributing degree-one \acp{VN} evenly across all \acp{CN} during \ac{PCM} construction optimizes the conditions for no \acp{4SET} to form.
This happens inherently if \ac{PEG} is used to construct the \ac{PCM}.
Numerical simulations support our theoretical results and demonstrate that distributing degree-one \acp{VN} evenly across all \acp{CN} using \ac{PEG} reduces the average \ac{PUPE} by a factor of $9.07$ compared to \ac{RCC}.
Furthermore, \ac{PEG} with the removal of all degree-one \ac{VN} stopping sets by employing the developed algorithm reduces the average \ac{PUPE} by a factor of $84.2$, compared to \ac{RCC}.

\bibliographystyle{IEEEtran}
\bibliography{IEEEabrv,bibliography}

\clearpage
\appendices

\section{Proof of Theorem~\ref{theorem:minimize-d}}\label{apx:proof-min-d}

First, we show that an uneven distribution of degree-one \acp{VN} across \acp{CN} can not minimize $|\mathcal{D}|$.
Then, we demonstrate that $|\mathcal{D}|$ is equal for all even distributions of \acp{VN} across \acp{CN}.

\begin{figure}[tbp]
  \centering
  \begin{tikzpicture}[%
  every node/.style={draw=black, fill=gray, minimum size = 2.5mm, inner sep = 0},
  vn/.style={circle},
  cn/.style={rectangle, minimum size = 1em, inner sep = .1em},
  dots/.style={
    draw=none,
    fill=none,
    anchor = center,
  },
]

  \node[cn] (cn1a) at (0,0) {\tiny$j_1$};
  \node[dots, below = of cn1a] (dots) {$\cdot\cdot\cdot$};
  \node[vn, left = 0.2em of dots] (vn-left) {};
  \node[vn, right = 0.2em of dots] (vn-right1) {};
  \node[vn, right = 0.2em of vn-right1] (vn-right2) {};
  \node[vn, right = 0.2em of vn-right2] (vn-right3) {};

  \draw (cn1a) -- (vn-left);
  \draw (cn1a) -- (vn-right1);
  \draw (cn1a) -- (vn-right2);
  \draw[KITred, thick] (cn1a) -- (vn-right3);

  \node[vn, right = 0.6em of vn-right3] (vn2-left) {};
  \node[dots, right = 0.2em of vn2-left] (dots2) {$\cdot\cdot\cdot$};
  \node[cn, above = of dots2] (cn2a) {\tiny$j_2$};
  \node[vn, right = 0.2em of dots2] (vn2-right) {};

  \draw (cn2a) -- (vn2-left);
  \draw (cn2a) -- (vn2-right);

  \node[cn] (cn1b) at (4,0) {\tiny$j_1$};
  \node[dots, below = of cn1b] (dotsb) {$\cdot\cdot\cdot$};
  \node[vn, left = 0.2em of dotsb] (vnb-left) {};
  \node[vn, right = 0.2em of dotsb] (vnb-right1) {};
  \node[vn, right = 0.2em of vnb-right1] (vnb-right2) {};
  \node[vn, right = 0.2em of vnb-right2] (vnb-right3) {};

  \draw (cn1b) -- (vnb-left);
  \draw (cn1b) -- (vnb-right1);
  \draw (cn1b) -- (vnb-right2);

  \node[vn, right = 0.6em of vnb-right3] (vn2b-left) {};
  \node[dots, right = 0.2em of vn2b-left] (dots2b) {$\cdot\cdot\cdot$};
  \node[cn, above = of dots2b] (cn2b) {\tiny$j_2$};
  \node[vn, right = 0.2em of dots2b] (vn2b-right) {};

  \draw (cn2b) -- (vn2b-left);
  \draw (cn2b) -- (vn2b-right);
  \draw[KITgreen, thick] (cn2b) -- (vnb-right3);

  \node[dots] (arr) at ({$0.5*(vn2-right)+0.5*(vnb-left)$} |- {$0.5*(cn1b)+0.5*(vn-left)$}) {$\to$};

\end{tikzpicture}
  \caption{Operation modifying one edge to spread degree-one \acp{VN} more evenly across \acp{CN}.}
  \label{apx:fig:proof-operation}
\end{figure}
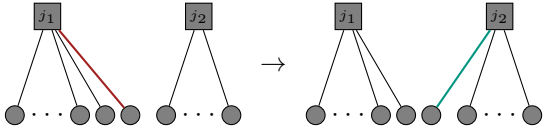

Assume a Tanner graph in which we distribute the degree-one \acp{VN} unevenly across all \acp{CN}.
This necessarily leads to two sets $\mathcal{V}_{j_1}$ and $\mathcal{V}_{j_2}$ with 
\begin{equation}
 |\mathcal{V}_{j_1}| \geq |\mathcal{V}_{j_2}| + 2.
 \label{apx:eq:assumption-vn-sets}
\end{equation}
Now, we consider the following operation, which is illustrated in Fig.~\ref{apx:fig:proof-operation}:
remove one degree-one \ac{VN} from \ac{CN} $j_1$ and move it to \ac{CN} $j_2$.
This results in
\begin{align}
  |\mathcal{V}_{j_1}'| &= |\mathcal{V}_{j_1}| - 1 \quad \text{and} \label{apx:eq:vj1}\\
  |\mathcal{V}_{j_2}'| &= |\mathcal{V}_{j_2}| + 1. \label{apx:eq:vj2}
\end{align}
Let the multisets $\mathcal{D}_{j_1}$, $\mathcal{D}_{j_2}$, $\mathcal{D}_{j_1}'$ and $\mathcal{D}_{j_2}'$ be the distance multisets corresponding to $\mathcal{V}_{j_1}$, $\mathcal{V}_{j_2}$, $\mathcal{V}_{j_1}'$ and $\mathcal{V}_{j_2}'$, respectively.
Then, with
\begin{equation}
  |\mathcal{D}_j| = \binom{|\mathcal{V}_j|}{2},
  \label{apx:eq:distance-set-size}
\end{equation}
we can show, that
\begin{equation}
  |\mathcal{D}_{j_1}| + |\mathcal{D}_{j_2}| > |\mathcal{D}_{j_1}'| + |\mathcal{D}_{j_2}'|.
  \label{apx:eq:decrease-in-d-set-size}
\end{equation}
Using~\eqref{apx:eq:distance-set-size} as well as~\eqref{apx:eq:vj1} and~\eqref{apx:eq:vj2}, we expand~\eqref{apx:eq:decrease-in-d-set-size} to
\begin{equation}
  {|\mathcal{V}_{j_1}| \choose 2} + {|\mathcal{V}_{j_2}| \choose 2} > {|\mathcal{V}_{j_1}| - 1 \choose 2} + {|\mathcal{V}_{j_2}| + 1 \choose 2},
\end{equation}
which we subsequently simplify using the definition of the binomial coefficient.
Finally, the inequality reduces to
\begin{equation}
  |\mathcal{V}_{j_1}| > 1 + |\mathcal{V}_{j_2}|,
\end{equation}
which is true, as it is equivalent to~\eqref{apx:eq:assumption-vn-sets}.
Thus, the statement~\eqref{apx:eq:decrease-in-d-set-size} is also true.

As $|\mathcal{D}| = \sum_j |\mathcal{D}_j|$, and only $\mathcal{D}_{j_1}$ and $\mathcal{D}_{j_2}$ are modified, the described operation reduces the cardinality~$|\mathcal{D}|$ of the total distance multiset.
Hence, any Tanner graph in which the degree-one \acp{VN} are not distributed evenly across all \acp{CN} does not minimize~$|\mathcal{D}|$.
Finally, in Tanner graphs in which they are distributed evenly, no operation can change the number of sets $\mathcal{V}_j$ with a given cardinality, without violating the even distribution.
Therefore, all Tanner graphs with an even distribution of degree-one \acp{VN} across \acp{CN} lead to the same~$|\mathcal{D}|$.

\section{Deterministic Construction for $V\leq2$}\label{apx:v2pcm}

We can construct a \ac{4SET}-free \ac{PCM} for an ensemble with $L_1 = 1$ and $r_\mathrm{d}=\frac{1}{2}$ by choosing $\mathcal{V}_j = \left\{ \frac{n+1}{2} \pm \left(j - \frac{1}{2}\right)\right\}$.
To show how this can be extended to other ensembles with $V\leq 2$, we outline how to convert a \ac{PCM} with no more than two degree-one \acp{VN} connected to any \ac{CN}, into a \ac{4SET}-free \ac{PCM}.
First, we identify all \acp{CN} connected to two degree-one \acp{VN} and select all degree-one \acp{VN} connected to them.
Then, we permute the \ac{PCM} columns corresponding to these \acp{VN} to resemble the structure described for $L_1=1$, $r_\mathrm{d}=\frac{1}{2}$.
The resulting \ac{PCM} is \ac{4SET}-free regardless of how the remaining \ac{PCM} columns are arranged.
For instance, the \ac{PCM}
\begin{equation}
  \bm{H} = 
  \begin{pmatrix}
    0 & 0 & 1 & 0 & 0 & 0 & 0 & 0 & 0\\
    1 & 0 & 0 & 0 & 1 & 0 & 0 & 1 & 0\\
    0 & 1 & 0 & 0 & 0 & 0 & 1 & 0 & 0\\
    0 & 0 & 1 & 1 & 0 & 0 & 0 & 1 & 1\\
    0 & 0 & 0 & 0 & 0 & 1 & 0 & 0 & 0
  \end{pmatrix}
\end{equation}
contains no more than two degee-one \acp{VN} per \ac{CN}.
The rows $2$, $3$ and $4$ correspond to \acp{CN} connected to two degree-one \acp{VN}, which correspond to columns $1$ and $5$, $2$ and $7$, as well as $5$ and $9$.
We permute these columns to form the structure found for $L_1=1$, $r_\mathrm{d}=\frac{1}{2}$:
\begin{equation}
  \begin{pmatrix}
    \color{lightgray} 0 & \color{lightgray} 0 & \color{lightgray} 0 & \color{lightgray} 0 & \color{lightgray} 0 & \color{lightgray} 0\\
    0 & 0 & 1 & 1 & 0 & 0\\
    0 & 1 & 0 & 0 & 1 & 0 & \cdots\\
    1 & 0 & 0 & 0 & 0 & 1\\
    \color{lightgray} 0 & \color{lightgray} 0 & \color{lightgray} 0 & \color{lightgray} 0 & \color{lightgray} 0 & \color{lightgray} 0
  \end{pmatrix}
\end{equation}
Finally, we append the remaining columns in an arbitrary order to form
\begin{equation}
  \bm{H}_\text{4SET-free} = 
  \begin{pmatrix}
    0 & 0 & 0 & 0 & 0 & 0 & 1 & 0 & 0\\
    0 & 0 & 1 & 1 & 0 & 0 & 0 & 0 & 1\\
    0 & 1 & 0 & 0 & 1 & 0 & 0 & 0 & 0\\
    1 & 0 & 0 & 0 & 0 & 1 & 1 & 0 & 1\\
    0 & 0 & 0 & 0 & 0 & 0 & 0 & 1 & 0
  \end{pmatrix}.
\end{equation}
\vspace{.3cm}

\section{Lower Bound on Distance Set Size}\label{apx:lagrange}

To lower bound the total distance multiset size
\begin{equation}
  |\mathcal{D}| = \sum_{j=1}^m \frac{|\mathcal{V}_j|(|\mathcal{V}_j| - 1)}{2}
  \label{apx:eq:cardinality-d}
\end{equation}
we use the fact that $f(x) = x(x-1)/2$ is convex.
Applying Jensen's inequality we can state
\begin{equation}
  \sum_{j=1}^m \frac{1}{m} f(|\mathcal{V}_j|) \geq f\left(\sum_{j=1}^m \frac{1}{m} |\mathcal{V}_j|\right) = f(V),
\end{equation}
where $V$ is the average number of degree-one \acp{VN} per \ac{CN}.
Multiplying both sides of the inequality with $m$ and applying~\eqref{apx:eq:cardinality-d} results in~\eqref{eq:bound-on-cardinality-d}.

\section{Test for Degree-One VN Stopping-Sets}\label{apx:finding-deg-one-vn-ss}

To check whether degree-one \ac{VN} stopping sets exist for a given delay $\tau$ and a given \ac{PCM} $\bm{H}$, we use the modified \ac{PCM} $\bm{H}^1 = (\bm{h}_1^1 \cdots \bm{h}_n^1)$, in which all columns not corresponding to degree-one \acp{VN} are set to contain only zeros.
This modified \ac{PCM} is then stacked to form
\begin{equation}
  \tilde{\bm{H}}^1 = 
  \begin{pmatrix}
    \bm{h}_\tau^1 & \cdots & \bm{h}_n^1\\
    \bm{h}_1^1    & \cdots & \bm{h}_{n-\tau}^1
  \end{pmatrix}
  \in \mathbb{F}^{2m \times (n-\tau)}.
\end{equation}
Subsequently, rows or columns of $\tilde{\bm{H}}^1$ with weight $\leq 1$ are removed iteratively.
If all entries of $\tilde{\bm{H}}^1$ can be removed in this way, $\bm{H}$ forms no degree-one \ac{VN} stopping set for the delay~$\tau$.

\end{document}